\documentclass[10pt,journal,compsoc]{IEEEtran}

\usepackage[T1]{fontenc}
\usepackage[utf8]{inputenc}
\usepackage{graphicx}
\usepackage{amsmath,amssymb}
\usepackage{textcomp}
\usepackage{microtype}
\usepackage[nocompress]{cite}
\usepackage[svgnames]{xcolor}
\usepackage{booktabs}
\usepackage{stfloats}
\usepackage{afterpage}
\usepackage{wrapfig}
\usepackage[inline]{enumitem}
\usepackage[hidelinks,bookmarksnumbered]{hyperref}
\usepackage[all]{hypcap}
\usepackage[nameinlink,capitalise]{cleveref}

\graphicspath{{figs/}}
\newcommand{\system}{FigFlux}

\title{Presenting Your Figures: Data-informed Animated Transitions Across Scientific Figure Panels}
\author{Lu~Ying, Can~Liu, Yuxiao~Yang, Jianing~Yin, Xiaolin~Wen, and Yong~Wang%
\IEEEcompsocitemizethanks{%
\IEEEcompsocthanksitem Lu Ying is with the Faculty of Computer Science and Data Science, University of Vienna, Vienna, Austria.
\IEEEcompsocthanksitem Yuxiao Yang and Jianing Yin are with the State Key Lab of CAD\&CG, Zhejiang University, Hangzhou, Zhejiang, China. E-mail: \href{mailto:yinjianing@zju.edu.cn}{yinjianing@zju.edu.cn} (Jianing Yin).
\IEEEcompsocthanksitem Can Liu, Xiaolin Wen and Yong Wang is with the College of Computing and Data Science, Nanyang Technological University, Singapore, Singapore.
\vspace{\baselineskip}}%
}
\hypersetup{
  pdftitle={Presenting Your Figures: Data-informed Animated Transitions Across Scientific Figure Panels},
  pdfauthor={Lu Ying, Can Liu, Yuxiao Yang, Jianing Yin, Xiaolin Wen, Yong Wang},
  pdfsubject={Manuscript submitted to IEEE Transactions on Visualization and Computer Graphics},
  pdfkeywords={Animated transitions, AI agents, multi-panel figures}
}

\IEEEtitleabstractindextext{%
\begin{abstract}
Multi-panel figures are commonly used in research papers to show scientific data, and researchers often reuse them when presenting their work.
Presenting these figures often involves explaining how data across chart panels are related.
Animated transitions can progressively reveal these relationships, but two static panels specify only the endpoints, leaving what intermediate keyframes should show and how the transition should progress unclear.
To support transition design across diverse scientific figures, we developed \emph{Figure Space} and \emph{Transition Space} through iterative analysis of 48 published multi-panel figures across eight scientific fields.
Figure Space represents panel composition, data transformations, and visual encodings while retaining the links needed to trace panel results back to records.
Transition Space uses this representation and source data to construct and sequence keyframes that make corresponding data visible and align them while preserving the meaning of each panel’s position encodings.
Building on these spaces, we developed FigFlux, a human-in-the-loop authoring system that guides AI agents in generating animated transitions from a multi-panel figure and its associated materials.
Researchers can revise keyframes in natural language before animation generation.
We evaluate FigFlux through cross-domain case studies, a study with 16 scientific authors, and expert assessment.
Authors reported lower overall workload than with their familiar tools, while experts gave FigFlux transitions higher ratings than transitions created with participants’ familiar tools or generated by AI alone. The example gallery have been published online: \href{https://figflux.github.io}{https://figflux.github.io}.
\end{abstract}
}

\AtBeginDocument{%
}
\crefname{figure}{Fig.}{Figs.}
\crefname{section}{Sec.}{Secs.}

\begin{document}
\maketitle
\IEEEdisplaynontitleabstractindextext

\begin{figure*}[!t]

  \centering
  \includegraphics[width=\linewidth, alt={Two FigFlux transition examples. The first row connects weekly temporal patterns to spatial partitions. The second row connects maps of the 2021 Pacific Northwest heatwave to a long-term temperature anomaly chart.}]{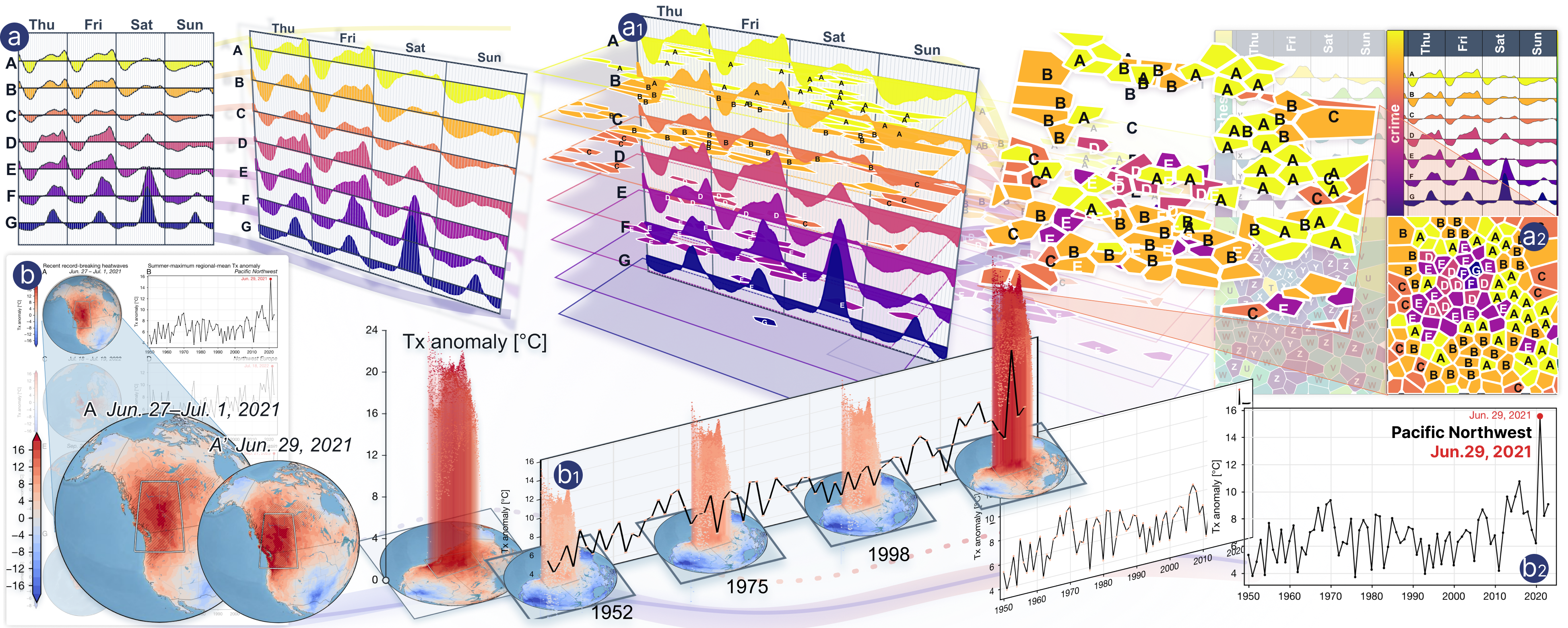}
  \caption{%
    Examples of \system{}-generated transitions using 3D Joint Coordinate Systems.
    (a) Weekly crime patterns: (\(a_1\)) an intermediate keyframe aligning each category's regions with its weekly curve, and (\(a_2\)) the final spatial partition map~\cite{prietocuriel2021heartbeat}.
    (b) The Pacific Northwest heatwave: (\(b_1\)) an intermediate keyframe aligning maps and temperature observations with the historical curve by year, and (\(b_2\)) the final line chart highlighting 2021~\cite{kornhuber2024global}.
  }
  \label{fig:teaser}
\end{figure*}

\section{Introduction}
\label{sec:introduction}

Multi-panel figures are common in scientific research papers across different fields.
For example, they account for roughly 30\% of figures 
in a large PubMed Central corpus~\cite{lee2018viziometrics}, and their proportion reaches 75\% in a biomedical literature dataset~\cite{subramanian2020medicat}.
Many such figures include chart panels that present research data~\cite{lee2018viziometrics} and have been widely used in research papers to support key findings~\cite{weissgerber2015beyond}.
Researchers use such figures to explain their findings and showcase their work.
For example, they may adapt these figures for academic presentations in general, or convert them into interactive formats for interactive research publications (e.g., Distill\footnote{\url{https://distill.pub/}}).

%

Presenting a multi-panel figure requires explaining how its chart panels jointly support the research findings.
Even when the audience understands each panel's result, they still need to know how data across panels relate.
For example, the source figure in \autoref{fig:teaser}(b) uses maps to show the spatial patterns of heatwaves and line charts to compare their intensities with previous years~\cite{kornhuber2024global}.
For each day, maximum temperature anomalies are averaged across the map’s boxed region, and the line chart highlights only the highest of these daily regional means in summer 2021.
In a paper, readers can examine these connections 
by cross-referencing
the panels, caption, and body text.
However,
when presenting a figure, researchers need to explicitly explain these data relationships as they move from one panel to the next.
%

Animated transitions can progressively reveal data relationships across panels, helping the audience connect the results shown in different chart panels~\cite{heer2007animated,tang2020narrative}.
However, two static panels specify only the start and end states of a transition.
What the intermediate keyframes should show and how the transition should progress remain unclear, particularly for scientific figures whose panels use different data transformations and visual encodings.
Supporting such transitions across diverse scientific figures poses two challenges: figure representation and keyframe design.
%

\begin{itemize}[noitemsep,topsep=0pt,parsep=0pt,partopsep=0pt, leftmargin=*,label=\textbullet]
\item \textbf{How to establish a general representation of multi-panel figures for transition design.}
Scientific figures vary in panel composition, data sources, and data transformations.
The data used to draw a panel may no longer retain the information needed to trace its results back to source records.
For example, connecting the summer peak in temperature anomalies to the map in \autoref{fig:teaser}(b) requires identifying the day of the peak and the temperature anomalies at locations within the boxed region that contributed to that day’s regional mean.
Existing visualization grammars, such as Vega-Lite, provide a foundation for describing data transformations, visual encodings, and view composition~\cite{satyanarayan2017vegalite}.
For transition design, these descriptions need to be organized across panels to identify corresponding data fields and compare how they are transformed and encoded.
The challenge is to represent diverse panel compositions consistently while retaining the information needed to connect displayed values to the source data used to compute them.
%

\item \textbf{How to construct and sequence transition keyframes that explain data relationships between chart panels.}
Although data correspondences can be identified from the source data, animated transitions must make these relationships explicit through changes that viewers can follow.
For example, in \autoref{fig:teaser}(a), a line shows a category’s weekly pattern, while a map shows the regions belonging to that category.
When transitioning from the line chart to the map, the line chart’s temporal and value axes need to be preserved.
When transitioning from the map to the line chart, the geographic layout of the regions needs to be preserved.
Previous work, including Gemini$^2$ and AniVis, supports intermediate keyframe recommendation or staged transition generation~\cite{kim2021gemini2,li2021anivis}.
However, these approaches rely on explicit chart specifications and primarily support only basic statistical charts.
Supporting panel pairs such as the line chart and map requires intermediate keyframes that accommodate their different coordinate structures while making their data correspondences explicit.
%

\end{itemize}

To address these challenges, we developed \emph{Figure Space} and \emph{Transition Space} through iterative analysis of 48 published multi-panel figures across eight scientific fields (e.g., physics, chemistry, and biology).
Figure Space introduces a \textit{Layout} and a \textit{Figure Table} to provide a unified representation that captures figure composition and retains the information needed to trace the data shown in each chart panel back to its source records.
The Layout describes panel juxtaposition, repetition, and overlay through composable operators, while the Figure Table aligns data fields across panels and retains each panel’s transformation steps and visual encodings.
%
%
Then, Transition Space uses this unified representation and source data to establish data correspondences and specify the data and visual encodings of intermediate keyframes, as well as their order.
When corresponding data cannot be identified separately in an original panel, we adjust its data transformations and introduce keyframes that show those data as distinct marks or mark groups.
We construct a two- or three-dimensional \emph{Joint Coordinate System} that aligns corresponding data while preserving the meaning of each panel’s position encodings.
We start with the target panel’s axes and reuse them for the source panel when possible.
We add a third axis for data that need their own scale, or place the source panel on a separate plane to preserve its spatial layout.
%

Built upon Figure Space and Transition Space, we developed \system{}, an authoring system that allows users to collaborate with AI agents to easily generate animated transitions across scientific figure panels while maintaining data correspondence.
Researchers can revise keyframes in natural language before confirming them for animation generation.
In a study with 16 scientific authors, \system{} received significantly higher ratings on several measures of authoring and explanation and produced a significantly lower overall workload score than participants’ familiar tools.
Four visualization experts also rated its transitions higher on average than those created with participants’ familiar tools or AI alone across all five criteria.
%

Our contributions are as follows:
\begin{itemize}[noitemsep,topsep=0pt,parsep=0pt,partopsep=0pt, leftmargin=*,label=\textbullet]
  \item \textbf{Figure Space}, a general representation of multi-panel scientific figures that captures panel composition and retains the information needed to trace panel results to source records.
  \item \textbf{Transition Space}, a design space for constructing and sequencing keyframes of an animated transition to explain data relationships between chart panels.
  \item \textbf{\system{}}, a human-in-the-loop authoring system that uses both spaces to guide AI agents in generating animated transitions. 
  We evaluate the system through cross-domain case studies, a comparative study with scientific authors, and expert assessment. 
\end{itemize}

\section{Related Work}
\label{sec:related-work}


\subsection{Multi-panel Figure Design}
\label{sec:related-work-multiview}

The design of scientific figures shapes how readers perceive and interpret quantitative information~\cite{cleveland1984graphical}.
Prior work recommends clear visual encodings, readable labels and annotations, informative captions, and consistent scales and styles across panels~\cite{wong2010design,rougier2014ten,wilke2019fundamentals,jambor2021creating}.
In a multi-panel figure, the graphs, plots, and diagrams should all address the same question and contribute to a clear take-home message~\cite{disis2024effective}.
Together, these studies provide general guidance for designing and organizing multi-panel scientific figures.
%

To examine chart panels more closely, we draw on research on multiple-view visualization, which studies how views are arranged and how data is related across them~\cite{baldonado2000guidelines,roberts2007state}.
Javed and Elmqvist identify five ways to combine views: juxtaposition, superimposition, overloading, nesting, and integration~\cite{javed2012exploring}.
Chen et al.\ analyze published visualizations to identify which view types appear together and how they are arranged~\cite{chen2021composition}.
Deng et al.\ identify eight recurring patterns based on where views are placed and how their contents are related~\cite{deng2023revisiting}.
For comparison, views can be placed side by side, overlaid, or designed to show their differences directly~\cite{gleicher2011visual,lyi2021comparative}.
Other work helps authors keep scales and value orders consistent across views and warns when visual encodings may suggest misleading relationships~\cite{qu2018keeping,kristiansen2022semantic}.
%

Another line of work examines relationships between data items shown in different views~\cite{javed2012exploring}.
Snap-Together coordinates selection and navigation based on joins between data tables~\cite{north2000snap}.
VisLink and ConnectedCharts draw links between related items, while View Relations explores designs for showing relationships between views~\cite{collins2007vislink,viau2012connectedcharts,knudsen2016view}.
SightBi computes groups of related items across views as biclusters and displays their relationships in separate views~\cite{sun2022sightbi}.
Sun et al.\ compare designs for showing cross-view data relationships and discuss their advantages and limitations~\cite{sun2022towards}.
These studies and the layout research above guide how we connect chart panels in published scientific figures.
We use panel layout to recommend pairs and shared data values to match marks or mark groups when constructing animated transitions.
%

\subsection{Animated Transitions and Authoring}
\label{sec:related-work-animation}

Animation is commonly used in visualization to show changes over time and transitions between views~\cite{fisher2010animation}.
It can help viewers follow changes in visual elements~\cite{fisher2010animation}.
Tversky et al.\ emphasized that animations should present information in ways that viewers can readily perceive and understand~\cite{tversky2002animation}.
For animated transitions, this involves deciding which changes to show together and how visual elements move.
Studies have also examined dividing changes into stages, starting elements at different times, adjusting the pace of movement, and designing motion paths~\cite{heer2007animated,chevalier2014not,dragicevic2011temporal,du2015trajectory,wang2017vector}.
%

Beyond these general animation techniques, constructing a visualization transition also depends on how data and visual elements correspond.
With clear correspondences, designers often animate matched elements by interpolating their positions, shapes, or visual encodings~\cite{heer2007animated,riche2007nodetrix}.
When such direct correspondence is insufficient to explain a change, transitions may divide the transformation into stages or introduce an intermediate representation.
Kim et al.\ used staged transitions to show how individual records are combined into aggregate values~\cite{kim2019aggregate}.
Tilt Map uses a prism map as an intermediate representation between choropleth maps and bar charts~\cite{yang2021tilt}.
Kiesel et al.\ introduced polycurve star plots as an intermediate representation between parallel coordinates and scatterplots~\cite{kiesel2023smooth}.
Salako et al.\ used staged transitions to connect node-link diagrams and parallel coordinates that emphasize different facets of the same data~\cite{salako2025animated}.
In data videos, Tang et al.\ describe preserving guides that reuse visual elements to connect scenes and narrative agents that represent data values or attributes during transitions~\cite{tang2020narrative}.
We draw on these studies to match marks across scientific chart panels using shared data values and to construct Joint Coordinate Systems based on each panel's data transformations.

Beyond studying how visualization transitions should be designed, researchers have developed tools that automate parts of the authoring process.
GraphScape models chart states and editing operations as a graph to recommend visualization sequences and paths between charts~\cite{kim2017graphscape}.
Canis and CAST enable authors to specify animated elements and timing through a declarative language or an interactive interface, respectively~\cite{ge2020canis,ge2021cast}.
Data Animator matches corresponding elements between two charts and generates a transition that authors can refine~\cite{thompson2021dataanimator}.
Gemini enumerates and ranks staged transition designs for author-specified start and end charts, while Gemini$^2$ recommends intermediate keyframes~\cite{kim2021gemini,kim2021gemini2}.
%

These tools generally require inputs that explicitly describe chart elements and their associated data, such as Vega specifications or data-enriched SVGs.
More recent systems can derive chart information from SVGs.
Live Charts recovers chart components and data to generate animations with audio narration~\cite{ying2025livecharts}.
WonderFlow identifies chart components and lets authors link narration text to them, synchronizing animations with speech~\cite{wang2025wonderflow}.
These two systems focus on animating information within a chart, whereas we focus on transitions between chart panels in published scientific figures.
Despite this difference, we drew on the techniques and tools reviewed above to develop Figure Space and Transition Space and to design and generate transitions.
%

\section{Multi-panel Figure Analysis}
\label{sec:design-space-development}

To inform the development of Figure Space and Transition Space,
we conducted iterative analyses of multi-panel scientific figures in scientific research publications.
Figure Space defines how to represent a figure's panel layout and each chart panel's source records, data transformations, and visual encodings.
For an ordered pair of chart panels, Transition Space organizes valid keyframe sequences that explain the data relationship between them.
%

\subsection{Corpus Scope and Construction}
\label{sec:space-development-corpus}

We treat each figure as the unit of analysis and define a panel as a distinct visual unit presenting a chart, scientific image, or schematic diagram.
We use \emph{chart} broadly to refer to a data-driven visualization, such as a plot, map, or network.
We identify panels through their spatial separation, labels, and descriptions in the caption.
A labeled subfigure may contain multiple panels if it includes distinct visual units.
%

We focus on published scientific figures containing at least two panels, more than half of which are charts.
Each figure must also contain at least one pair of chart panels that show data for some of the same entities, categories, locations, or time periods.
Scientific images and schematic diagrams may appear as supporting panels.
We exclude figures dominated by domain-specific scientific imagery because animating such content requires specialized data and rendering pipelines.
%

We sampled articles from each of the eight top-level subject areas in the Nature Portfolio taxonomy.
For each area, we examined figures published between 2020 and 2025 in the Nature and Science journal families and leading journals in that field.
We prioritized articles that provided source data and plotting code to make their chart panels easier to reproduce.
When an article spanned multiple subject areas, we assigned it to one area based on its primary research question and counted it only once.
We selected one figure from each of 48 articles.
For each selected figure, we collected its image, caption, relevant passages from the article, and available supporting materials.
%

\subsection{Iterative Development Procedure}
\label{sec:characterizing-analysis}

Based on prior work on visualization grammars~\cite{satyanarayan2017vegalite}, composite visualizations, and multiple-view visualizations~\cite{javed2012exploring,chen2021composition,deng2023revisiting,sun2022sightbi,sun2022towards}, we first examined how each chart panel was produced from source data, how panels were arranged, and how their data related.
We developed and refined Figure Space and Transition Space in two analysis rounds.
%

In the first round, the three authors jointly selected 16 seed figures, two from each subject area.
We considered panel types appearing together, panel arrangement, and which records the chart panels showed and how they grouped or aggregated them.
%

All three authors independently analyzed the 16 seed figures.
For each figure, they described its panel layout and examined how source data were transformed and visually encoded in each chart panel.
They then selected panel pairs with related data and designed candidate transitions to explain those relationships.
For Figure Space, each author proposed rules for describing panel layouts and linking chart panels to source data.
For Transition Space, they proposed constraints on valid keyframe sequences between chart panels.
The authors met weekly to compare their proposals and revise the two spaces by consensus.
They reanalyzed the seed figures after each revision, repeating this process until all 16 figures could be represented in Figure Space and their candidate transitions specified in Transition Space.
%

In the second round, one author selected 32 additional figures, four from each subject area.
This author analyzed them using the first round's representation rules and keyframe constraints.
A second author checked each analysis against the figure, caption, and relevant article passages.
All three authors discussed cases that revealed gaps or ambiguities in the representation rules or keyframe constraints.
They revised both spaces as needed and reapplied each revision to earlier cases.
We finalized both spaces after resolving the remaining issues.
%

%

\section{Figure Space}
\label{sec:figure-space}

We define Figure Space as the set of multi-panel scientific figure representations specified by a \textbf{Layout} and a \textbf{Figure Table}.
The Layout describes how panels are combined, while the Figure Table records how source data are transformed and visually encoded to produce each chart panel.
With the source data, the Layout and Figure Table provide the information needed to examine data relationships between panels.
%

In this section and \autoref{sec:transition-space}, we use a published figure on water storage potential for irrigation~\cite{schmitt2022global} as a running example.
The figure combines a world map with two bar charts comparing selected river basins.
%

\subsection{Layout}
\label{sec:figure-space-layout}

We identify panels following the criteria in \autoref{sec:space-development-corpus}.
We name chart panels by chart type and non-chart panels by content type, retaining source labels when available.
%

\begin{figure}[!htbp]
  \centering
  \includegraphics[width=\columnwidth, alt={Layout operators and a composite example. (a) Two panels are juxtaposed. (b) A shared panel specification is repeated across data values. (c) One panel is overlaid on another. (d) Repeated pie-chart and heatmap groups are overlaid on a world map. A detail view shows the Central Europe group.}]{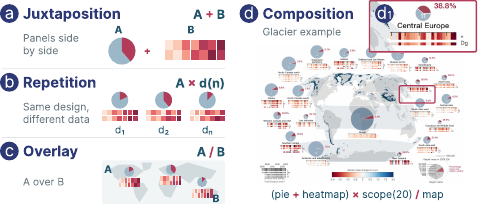}
  \caption{Layout operators and their composition. (a) Juxtaposition. (b) Repetition. (c) Overlay. (d) Composition of these operators in a glacier mass-change figure~\cite{glambie2025community}. (\(d_1\)) Detail of the pie chart and heatmap for Central Europe.}
  \label{fig:figure-layout}
\end{figure}

Drawing on prior work on composite and multiple-view visualizations~\cite{javed2012exploring,chen2021composition,deng2023revisiting,satyanarayan2017vegalite}, we describe panel organization using three composable operators, illustrated in \autoref{fig:figure-layout}(a--c).
Juxtaposition, \(A+B\), places panels or panel groups in separate regions.
Repetition uses the same chart design to show different groups of data.
We write \(A\times d(n)\) for \(n\) repetitions of panel or panel group \(A\), one for each selected value of data dimension \(d\).
Overlay, \(A/B\), places \(A\) over \(B\).
%

The glacier example pairs a pie chart with a heatmap for each scope, as the Central Europe detail shows (\autoref{fig:figure-layout}(\(d_1\)))~\cite{glambie2025community}.
This pair repeats across 20 scopes, comprising 19 glacier regions and the global estimate.
The repeated pairs overlay a map, giving the Layout in \autoref{fig:figure-layout}(d).
%

In our running example, two bar charts share a design for selected basin rank groups 1--5 and 6--30, with separate y-axis ranges.
The map is juxtaposed with these charts, giving the Layout \(\mathrm{map}(A)+\bigl(\mathrm{bar}(B)\times\mathrm{rankGroup}(2)\bigr)\), where \(A\) and \(B\) are the original panel labels.
%

\subsection{Figure Table}
\label{sec:figure-space-table}

To construct transitions between chart panels, we compare the data they show and how those data are visually encoded.
The Figure Table aligns fields across panels in shared rows, with \textbf{Transform} and \textbf{Encoding} columns for each panel.
Non-chart panels are omitted from the Figure Table.
For repeated panels, the table records one representative panel and the corresponding value of \(d\).
%

We describe the source data used by each chart panel as one or more tables~\cite{satyanarayan2017vegalite}.
Each column represents a field and specifies one kind of information recorded in the table.
The column header gives the field name.
Each row represents a record containing values for these fields.
Each cell contains the value of a particular field in a particular record.
Different panels may share a source table or use different source tables.
%

In our running example, each record describes a main river basin, with fields for its identifier, name, and total storage-fed irrigation \(F\).
%

\begin{table*}[!b]
  \caption{Figure Table for the irrigation map (A) and bar chart (B, ranks 1--5). Irrigation volumes are in \(\mathrm{km}^3/\mathrm{yr}\).}
  \label{tab:figure-table-example}
  \scriptsize
  \centering
  \setlength{\tabcolsep}{3pt}
  \renewcommand{\arraystretch}{1.0}
  \begin{tabular}{@{}p{0.07\textwidth}p{0.335\textwidth}p{0.19\textwidth}p{0.08\textwidth}p{0.19\textwidth}p{0.075\textwidth}@{}}
    \toprule
    Type & Dimension / Level / Measure & \multicolumn{2}{c}{map (A)} & \multicolumn{2}{c}{bar (B)} \\
    \cmidrule(lr){3-4}\cmidrule(lr){5-6}
    & & Transform & Encoding & Transform & Encoding \\
    \midrule
    dimension & space & --- & --- & --- & --- \\
    level & \(\hookrightarrow\) main river basin & \texttt{lookup[basin geometry]} & --- & \textcircled{\scriptsize 1} & \(\rightarrow x\) \(\cdot\) text \\
    measure & Total storage-fed irrigation (\(F\)) & --- & \(\rightarrow\) color & \textcircled{\scriptsize 1} & \(\rightarrow y\) \(\cdot\) color \\
    measure & Storage-fed irrigation on currently irrigated land (\(C\)) & --- & --- & \textcircled{\scriptsize 1} \(\rightarrow\) \texttt{filter[}\(C>0\)\texttt{]} & \(\rightarrow y\) \\
    measure & Total sustainable irrigation (\(S\)) & --- & --- & \textcircled{\scriptsize 1} \(\rightarrow\) \texttt{calculate[}\(I_{\mathrm{cur}}+I_{\mathrm{fut}}\)\texttt{]} & \(\rightarrow y\) \\
    \bottomrule
  \end{tabular}
  \par\smallskip
  \begin{minipage}{\textwidth}
    \textbf{Result \textcircled{\scriptsize 1}.} Among records with defined \(F\), select 30 basins by descending \(F-C\), order these basins by descending \(F\), and retain ranks 1--5.
    \(I_{\mathrm{cur}}\) denotes current irrigation volume, and \(I_{\mathrm{fut}}\) denotes additional irrigation potential on currently rainfed cropland.
  \end{minipage}
\end{table*}

Inspired by multidimensional data models~\cite{gray1997data,nguyen2000object}, we describe the data used by chart panels in terms of dimensions, levels, and measures.
A dimension describes an aspect used to organize and analyze data, such as space or time.
Levels organize values within a dimension at different granularities and may form one or more hierarchies.
For example, year, month, and day are levels of the time dimension.
A measure is a numerical field recorded directly or derived through an operation such as aggregation.
In our running example (\autoref{tab:figure-table-example}), we use space as a dimension and main river basin as a level, with individual basins as its values.
Basin ID identifies each basin, and Basin name provides its label.
Both fields are represented in the main river basin row of the Figure Table.
Total storage-fed irrigation is a measure.
%

The Transform column specifies operations on the data, such as filtering records, aggregating values, or deriving new fields.
The Encoding column specifies how fields in the resulting data map to visual channels, such as position, color, or size.
\autoref{tab:figure-table-example} presents a compact Figure Table for the map and the bar chart with \(\mathrm{rankGroup}=\mbox{1--5}\).
%

For Transform, we use operation names from Vega-Lite~\cite{satyanarayan2017vegalite}, such as filter, aggregate, bin, calculate, fold, lookup, and window.
To keep the Figure Table concise, we write each operation as \texttt{transform[parameters]}, where \texttt{transform} names the operation and brackets enclose its parameters.
Semicolons separate parameter groups within an operation.
An arrow (\(\rightarrow\)) connects two operations when the second uses the first's result.
For example, \texttt{filter[year = 2021]} \(\rightarrow\) \texttt{aggregate[mean; groupby=location]} first selects rows for 2021 and then computes the mean of the field identified by the table row for each location.
Transforms that independently read the same input use separate lines.
When the table row already identifies an input or output field, we omit the corresponding \texttt{field} or \texttt{as} property.
%

We label reusable transform results with circled identifiers, such as \textcircled{\scriptsize 1}.
In \autoref{tab:figure-table-example}, \textcircled{\scriptsize 1} represents the data for the five basins selected for bar (B).
The basin and \(F\) rows use \textcircled{\scriptsize 1} directly.
For the hatched bar regions, \mbox{\textcircled{\scriptsize 1} \(\rightarrow\) \texttt{filter[}\(C>0\)\texttt{]}} selects the basins with \(C>0\) from these five basins.
The notation \textcircled{\scriptsize 1}\texttt{.field} refers to a field in result \textcircled{\scriptsize 1}.
%

For Encoding, Vega-Lite's channel names~\cite{satyanarayan2017vegalite} describe how fields map to visual properties.
Because the row already identifies the field, each \textbf{Encoding} cell contains only \(\rightarrow\) \texttt{channel}.
A centered dot separates multiple channels.
For example, in \autoref{tab:figure-table-example}, bar (B) encodes total storage-fed irrigation (\(F\)) by bar height and color, written as \mbox{\(\rightarrow y\) \(\cdot\) color}.
A dash (---) marks cells with no transform or encoding.
%

In \autoref{tab:figure-table-example}, bar (B) shows five basins from map (A).
Both encode \(F\) by color, and bar (B) also uses height.
Only bar (B) shows \(C\) and \(S\).
%

\section{Transition Space}
\label{sec:transition-space}

\begin{figure*}[!t]
  \centering
  \includegraphics[width=\textwidth, alt={Data relationships and Joint Coordinate System. Panel a shows the irrigation map and bar chart, the supporting rows for their five shared basin IDs retained separately in Shared Data, and the additional basin rows included in Combined Data. Panel b illustrates reusing axes in b1, arranging marks in b2, adding a Z scale in b3, and preserving panel planes in b4, which also shows the irrigation example.}]{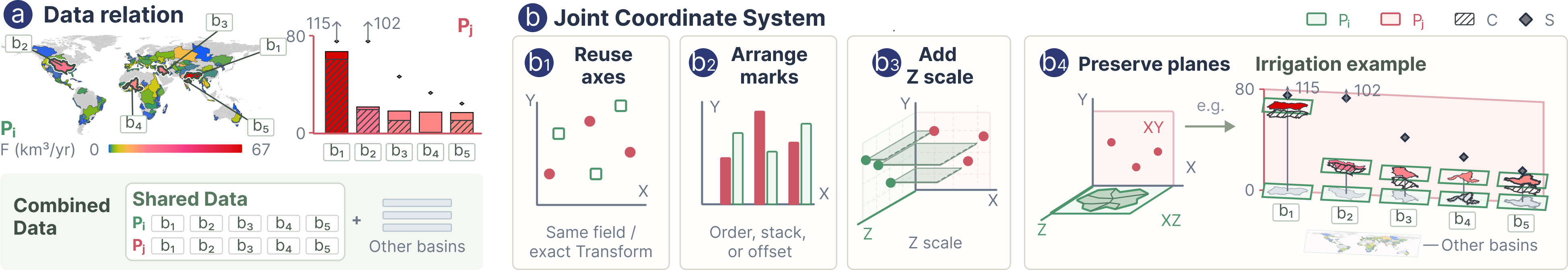}
  \caption{Data relationships and Joint Coordinate System. (a) Shared Data retains the supporting rows for the five basin IDs shared by the map \(P_i\) and bar chart \(P_j\), separately for each panel. Combined Data contains all source rows represented by either panel. (b) Joint Coordinate System: reusing axes (\(b_1\)), arranging marks (\(b_2\)), adding a \(Z\) scale (\(b_3\)), or preserving panel planes (\(b_4\)). The irrigation example in (\(b_4\)) preserves basin shapes on planes parallel to \(XZ\) and aligns them with the bars.}
  \label{fig:transition-space}
\end{figure*}

For an ordered pair of chart panels \(P_i\) and \(P_j\) represented in Figure Space, Transition Space describes the possible keyframe sequences that explain their data relationship.
Each keyframe specifies a visual state, including its data and Transforms, marks, visual encodings, and panel placement within a coordinate system.
Shared Data and Combined Data describe the underlying data, while Joint Coordinate System specifies coordinate arrangements compatible with the panels' encodings (\autoref{fig:transition-space}).
%

\subsection{Shared Data}
\label{sec:transition-space-shared-data}

The two panels may represent different subsets or aggregations of the source data.
To explain their relationship through keyframe sequences, we identify what they have in common and which source records in each panel support that connection.
%

Following prior approaches that establish correspondence between views~\cite{heer2007animated,thompson2021dataanimator}, we define Shared Data as the rows in each panel's source tables that support shared dimension-level values.
For each such level \(\ell\), we compute these values as
\[
  \mathrm{SharedValues}[\ell]
  =
  \mathrm{values}(P_i,\ell)
  \cap
  \mathrm{values}(P_j,\ell),
\]
where \(\mathrm{values}(P,\ell)\) denotes the values at level \(\ell\) supported by the source rows represented in panel \(P\).
For each shared value, Shared Data retains the supporting rows separately for the two panels.
The associated Transforms specify how the shared value is derived and how its supporting rows are selected.
The retained rows may differ between panels.
%

Measures alone do not establish data correspondence.
For each shared measure, its values are retained or computed separately from each panel's supporting rows in Shared Data.
Any required aggregation or calculation is recorded in the associated Transforms.
The resulting measure values need not be equal.
A panel pair has a valid data correspondence when it shares at least one dimension-level value and all associated Transforms can be executed from the source data.
%

In our example (\autoref{tab:figure-table-example} and \autoref{fig:transition-space}(a)), the five basin IDs represented in both the map and the bar chart are the shared dimension-level values.
Shared Data consists of the source rows supporting these IDs, retained separately for each panel.
The shared measure, total storage-fed irrigation (\(F\)), is already recorded for each basin, so its values are retained without additional aggregation or calculation.
%

\begin{figure*}[!b]
  \centering
  \includegraphics[width=\textwidth, alt={Keyframe sequences in Transition Space. Panel a shows seven possible keyframe types in sequence order, with rows for data and Transforms, mark types, encodings, coordinate arrangements, and inclusion conditions. Panels b1 through b4 show the irrigation example: the map, the map with Joint Encoding, Combined Data in Joint Coordinate System, and the bar chart. Dashed boxes explain why the adjusted panels and the bar with Joint Encoding are omitted.}]{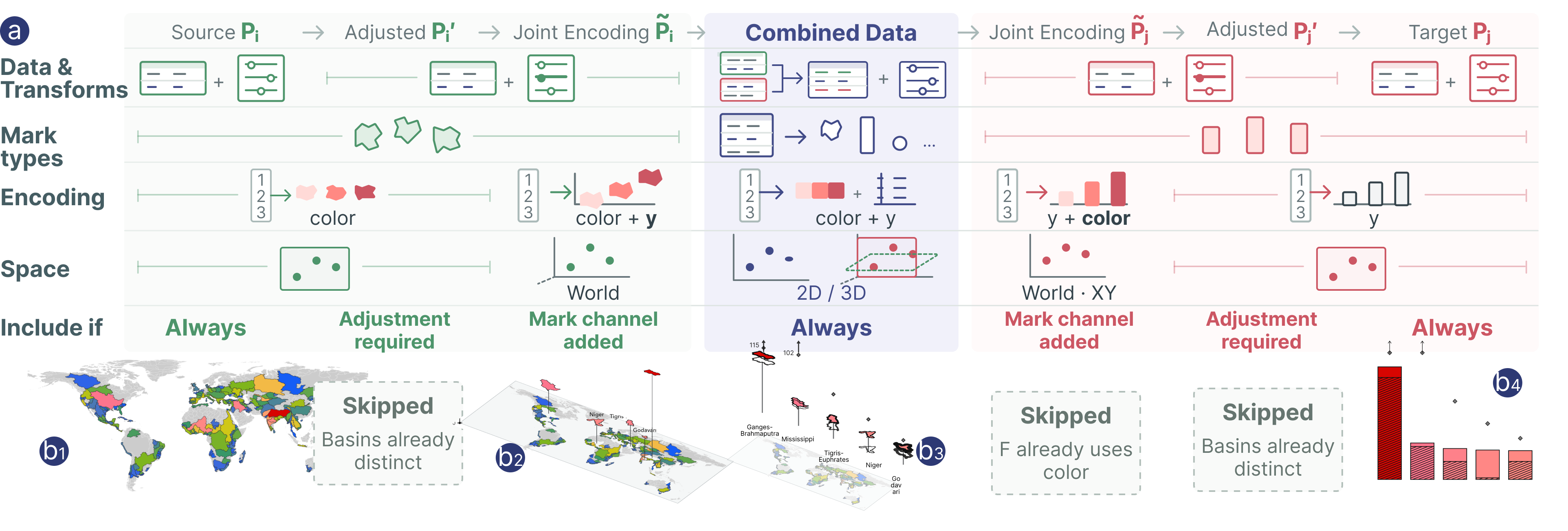}
  \caption{Keyframe sequences in Transition Space. (a) Keyframe types in sequence order, showing their components and when each keyframe is included. The Encoding row shows retained and added channels. (\(b_1\)--\(b_4\)) The irrigation example contains four keyframes: the map, the map with Joint Encoding, Combined Data in Joint Coordinate System, and the bar chart. Dashed boxes explain why the other keyframes are omitted.}
  \label{fig:keyframe-sequences}
\end{figure*}

\subsection{Combined Data}
\label{sec:transition-space-combined-data}

Combined Data contains the union of the source rows represented by the two panels after any necessary Transform adjustments, including the supporting rows in Shared Data and those represented by only one panel.
With the associated Transforms, it retains the information needed to recover the data represented by these panels.
%

To make the data correspondence visible, each shared dimension-level value must map to a separately identifiable mark or mark group in both panels.
If a panel does not meet this requirement, its Transforms are adjusted so that these values are represented separately.
Only the necessary Transform operations are modified, while other Transform branches, Encodings, and mark types remain unchanged.
An adjusted panel is denoted \(P_i'\) or \(P_j'\).
For example, a map showing multiday mean temperature anomalies can be filtered to the day highlighted in the target chart, retaining its geographic marks and color encoding (\autoref{fig:teaser}(b)).
%

\(\widetilde{P}_i\) denotes \(P_i\) when no adjustment is needed and \(P_i'\) otherwise, with \(\widetilde{P}_j\) defined in the same way.
Combined Data is given by
\[
  \mathrm{CombinedData}
  =
  \mathrm{rows}(\widetilde{P}_i)
  \cup
  \mathrm{rows}(\widetilde{P}_j),
\]
where \(\mathrm{rows}(P)\) denotes all source rows represented by panel \(P\), with each row identified by its source table and its identity within that table.
%

The associated Transforms derive dimension-level values from all rows in Combined Data and specify how its measures are obtained.
For each shared measure, any required aggregation or calculation specified for Shared Data is applied to the corresponding source rows in Combined Data.
In our irrigation example (\autoref{tab:figure-table-example}), Combined Data contains all basin rows represented by the map, including the five shared with the bar chart.
In the popularity example (\autoref{fig:case}(b)), each retained row records a participant's observed popularity and their estimate of it.
The associated Transforms count rows within each histogram bin and each pair of popularity values to obtain histogram and bubble frequencies, respectively.
%

\subsection{Joint Coordinate System}
\label{sec:transition-space-joint-space}

Joint Coordinate System is a common coordinate system for representing the data from both panels.
For each dimension-level or measure field \(f\) represented in Combined Data, Joint Encoding combines the visual channels used for that field in the two panels:
\[
  \mathrm{JointEncoding}[f]
  =
  \mathrm{Encoding}_{\widetilde{P}_i}[f]
  \cup
  \mathrm{Encoding}_{\widetilde{P}_j}[f].
\]
In \autoref{tab:figure-table-example}, \(F\) uses color in both panels and \(y\) in the bar chart, so \(\mathrm{JointEncoding}[F]=\{y,\mathrm{color}\}\).
The same Joint Encoding is used to represent Shared Data and Combined Data.
%

The target panel \(\widetilde{P}_j\) is the reference for the common coordinate system (World).
Its local \(x\)- and \(y\)-position channels map to World \(X\) and \(Y\), respectively.
In a three-dimensional Joint Coordinate System, \(Z\) is perpendicular to the \(XY\) plane (\autoref{fig:transition-space}(\(b_3\)--\(b_4\))).
%

Joint Coordinate System can be two- or three-dimensional.
A two-dimensional Joint Coordinate System positions corresponding marks within a single \(XY\) coordinate system, with at most one position scale per axis.
Marginal histograms may use separate length scales to encode frequency.
A position field can reuse an axis when it matches the field already encoded there or can be rewritten without loss through an exact Transform.
Otherwise, marks can be ordered, stacked, or offset, provided that these changes preserve the meaning of the original position encoding.
A three-dimensional Joint Coordinate System is used when these options cannot accommodate all position encodings.
%

In a three-dimensional Joint Coordinate System, \(\widetilde{P}_j\) stays on the \(XY\) plane.
A position field in \(\widetilde{P}_i\) shares an existing \(X\) or \(Y\) scale when possible and uses \(Z\) if it requires an independent scale.
If the original spatial arrangement of \(\widetilde{P}_i\) must be preserved, the panel occupies a plane parallel to \(XY\), \(XZ\), or \(YZ\).
%
In our irrigation example (\autoref{tab:figure-table-example}), the target bar chart uses \(X\) for basin order and \(Y\) for \(F\).
The five shared basins retain their geographic shapes on planes parallel to \(XZ\), at the same \(X\) and \(Y\) positions as the corresponding bars (\autoref{fig:transition-space}(\(b_4\))).
%

\subsection{Keyframe Sequences}
\label{sec:transition-space-keyframe-sequences}

A keyframe sequence starts with \(P_i\), includes a representation of Combined Data in the Joint Coordinate System using Joint Encoding, and ends with \(P_j\).
Shared dimension-level values and their supporting rows in Shared Data support correspondence between keyframes.
Additional keyframes represent panels with adjusted Transforms or added visual channels as needed to reveal this correspondence.
\autoref{fig:keyframe-sequences}(a) summarizes keyframe components and inclusion conditions.
%

The keyframes \(P_i'\) and \(P_j'\) represent panels with adjusted Transforms, retaining their original mark types, visual encodings, and coordinate systems.
An adjusted keyframe is included only when the original panel does not show each shared value as a distinct mark or mark group.
%

The keyframes \(\widetilde{P}_i\) and \(\widetilde{P}_j\) with Joint Encoding retain their panels' data, Transforms, and mark types while adding selected channels from the other panel for shared fields.
Each panel's local coordinate system is mapped into the Joint Coordinate System.
Each keyframe is included only when at least one channel is added to its marks.
In \autoref{fig:keyframe-sequences}(\(b_2\)), the five shared basins retain their geographic shapes and color encoding, with \(F\) also encoded by their positions along World \(Y\).
%

The Combined Data in Joint Coordinate System keyframe represents both panels' data using the associated Transforms and Joint Encoding.
Its marks and panels are arranged in the Joint Coordinate System, adjusting panel number or placement as needed.
In \autoref{fig:keyframe-sequences}(\(b_3\)), the shared basins retain these \(Y\) positions and follow the bar chart's order along World \(X\).
The remaining basins stay on a separate map plane.
%

Adjacent keyframes with identical components are merged.
In \autoref{tab:figure-table-example}, \(P_i'\) and \(P_j'\) are omitted because both panels already distinguish the shared basins.
The bar with Joint Encoding is also omitted because it already encodes \(F\) by color.
The sequence in \autoref{fig:keyframe-sequences}(\(b_1\)--\(b_4\)) contains four keyframes: the map, the map with Joint Encoding, Combined Data in Joint Coordinate System, and the bar chart.
%

\section{\system}
\label{sec:transition-generation-pipeline}

Based on Figure Space and Transition Space, we developed \system{}, an authoring system that guides AI agents in generating animated transitions between chart panels.
Users select the panels to connect and can revise the generated keyframes before confirming them for animation generation (\autoref{fig:system-overview}).
%

\subsection{System Overview}
\label{sec:transition-generation-overview}

After the user provides a figure and its associated data, source paper, and rendering code, \system{} identifies the panels and presents them for selection.
The user can select multiple panels for pairing or directly specify a source and target.
The Figure Analysis Agent analyzes how the selected panels are organized and how their data are transformed and visually encoded, producing a Layout and a Figure Table.
%

When multiple panels are selected, the Transition Pairing Agent uses the Layout to recommend panel pairs (\autoref{sec:transition-pairing}).
The user can select one or more of these pairs for transition generation.
Each selected or directly specified pair is processed independently.
%

\system{} identifies each panel pair's Shared Data (\autoref{sec:transition-space-shared-data}).
Processing stops if Shared Data cannot be obtained.
Otherwise, it constructs and renders a keyframe sequence for user review (\autoref{sec:keyframe-construction}).
%

\begin{figure}[htbp]
  \centering
  \includegraphics[width=\columnwidth, alt={The FigFlux interface at the keyframe representing Combined Data. A left sidebar lists the source figure and available transitions, a central workspace shows this keyframe and the motion timeline, and a right panel shows transition context and agent controls.}]{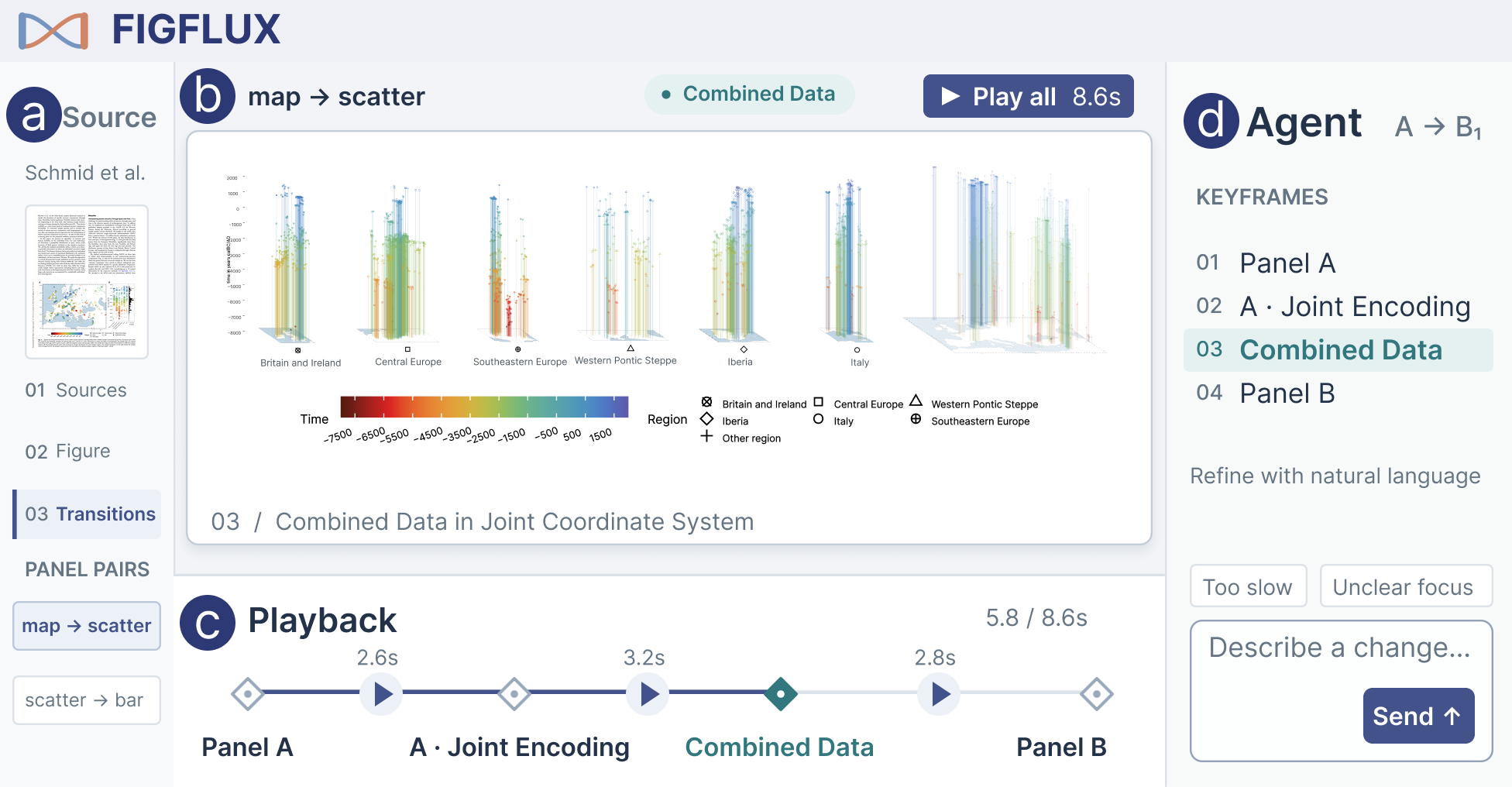}
  \caption{The \system{} interface. (a) Source information and panel-pair selection. (b) Animation preview at the Combined Data in Joint Coordinate System keyframe. (c) Playback controls and timeline. (d) Keyframe selection and editing.}
  \label{fig:system-overview}
\end{figure}

The user can request keyframe changes in natural language (\autoref{fig:system-overview}(d)).
\system{} updates and rerenders the keyframes accordingly.
After the user confirms them, the Transition Agent generates the animation (\autoref{sec:transition-generation}).
The Animation Refiner Agent then refines it to improve continuity and visual clarity.
The user can play or scrub the sequence and export the animation with its related files, including a standalone interactive player, an MP4 video, and editable transition source code.
%

\subsection{Transition Pairing}
\label{sec:transition-pairing}

The Transition Pairing Agent recommends panel pairs using the Layout operators (\autoref{sec:figure-space-layout}).
For \(A+B\), it orders the selected panels from left to right within each row and from top to bottom across rows, then recommends a transition from each panel to the next.
For \(A/B\), it keeps \(B\) fixed as the background and applies the Layout rules within \(A\) when at least two panels are selected there.
For \(A\times d(n)\), automatic pairing requires an order provided by the repeat dimension or the displayed arrangement.
The Agent recommends only the first adjacent pair in this order by default.
Users can directly specify the source and target to select other pairs or connect repeated panels without a defined order.
The Transforms for each selected repeated panel specify the value of \(d\) that identifies its data subset.
%
For nested Layouts, the Agent applies the rules within each group before determining connections between groups.
For example, when all panels in \(A+(B\times\mathrm{phase}(8))\) are selected, the repeated group yields \(B[\mathrm{phase}_1]\rightarrow B[\mathrm{phase}_2]\).
The outer juxtaposition connects \(A\) to the first selected panel in that group, yielding \(A\rightarrow B[\mathrm{phase}_1]\).
%

\subsection{Keyframe Construction}
\label{sec:keyframe-construction}

Given a panel pair selected by the user, \system{} uses six agents to construct and render a keyframe sequence.
The process has three stages: identifying Shared Data and adjusting panel Transforms when needed, constructing Combined Data and selecting a Joint Coordinate System, and specifying and rendering the keyframes.
%

\begin{figure}[htbp]
  \centering
  \includegraphics[width=\columnwidth, alt={Four rows showing correspondence between adjacent keyframes. Confirmed record-level correspondence preserves individual mark identity. A declared Transform maps n input marks to m output marks. When a shared level value is the only relation, marks representing that value are treated as groups. Without an available relation, source and target marks exit and enter independently.}]{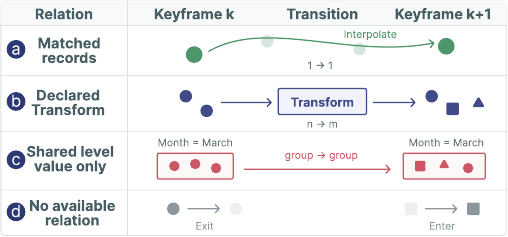}
  \caption{Different types of correspondence between adjacent keyframes. (a) Matched records. (b) Declared Transform. (c) Shared level value only. (d) No available relation.}
  \label{fig:transition-correspondence}
\end{figure}

\begin{figure*}[!b]
  \centering
  \includegraphics[width=\textwidth, alt={Two FigFlux case studies. The top row shows a transition from a histogram of higher-order network measures through an intermediate keyframe to a network. The bottom row shows a transition from popularity histograms through an intermediate keyframe to a bubble scatter plot. Both transitions use 2D Joint Coordinate Systems.}]{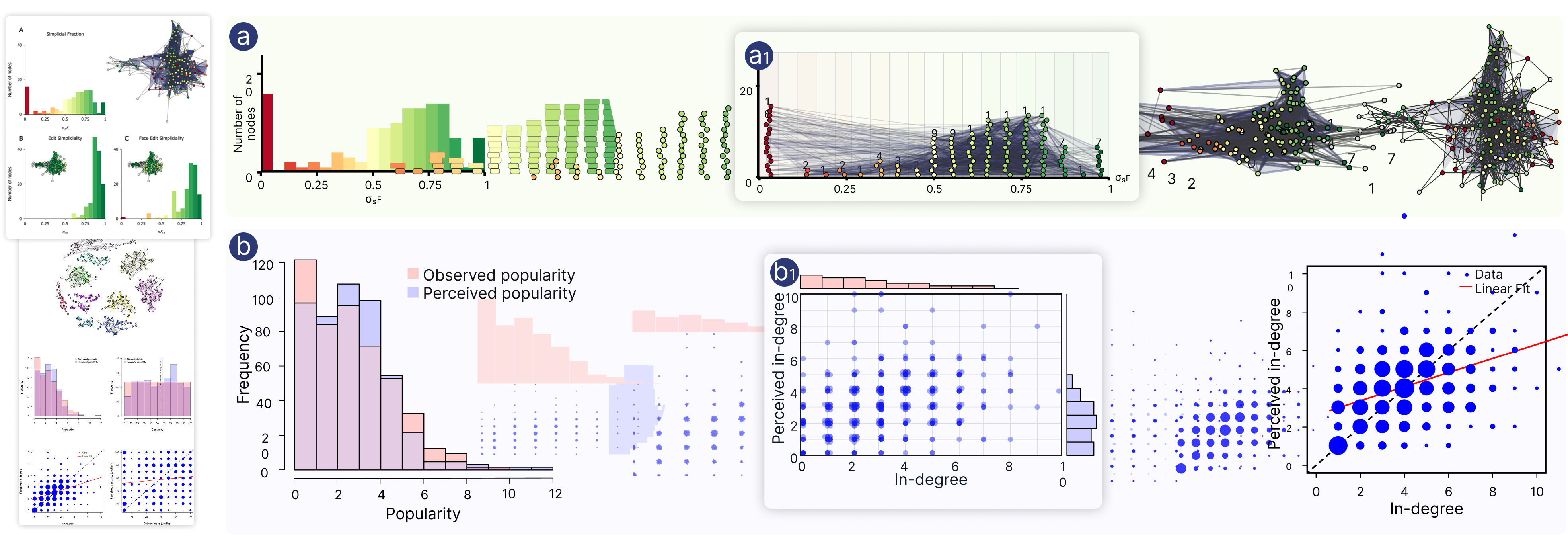}
  \caption{Examples of \system{}-generated transitions using 2D Joint Coordinate Systems.
  (a) Histogram to higher-order network: (\(a_1\)) an intermediate keyframe arranging nodes within histogram bins~\cite{landry2024simpliciality}.
  (b) Popularity histograms to bubble scatter plot: (\(b_1\)) an intermediate keyframe showing one mark per participant~\cite{kovarik2025perception}.}
  \label{fig:case}
\end{figure*}

In the first stage, the Shared Data Agent receives the panel pair, source data, and Figure Table and returns Shared Data and its associated Transforms (\autoref{sec:transition-space-shared-data}).
Processing stops if Shared Data cannot be obtained.
Otherwise, the Alignment Agent uses these outputs and the rendering code to apply any required Transform adjustments (\autoref{sec:transition-space-combined-data}).
The resulting panels, \(\widetilde{P}_i\) and \(\widetilde{P}_j\), are passed to the Combined Data Agent.
%

In the second stage, the Combined Data Agent takes Shared Data, \(\widetilde{P}_i\), and \(\widetilde{P}_j\) as input and produces Combined Data, its associated Transforms, and Joint Encoding.
The Joint Coordinate System Agent uses these outputs, the panels, source data, and rendering code to generate candidate configurations following \autoref{sec:transition-space-joint-space}.
It first attempts a two-dimensional Joint Coordinate System with a single panel and considers three-dimensional configurations only if no valid candidate is found.
Candidates are retained only if they represent all Combined Data and make every shared dimension-level value identifiable.
The remaining candidates are ranked by the clarity of data correspondence, preservation of the original panels, and spatial clarity, in that priority order.
The Agent returns all valid candidates as a ranked list.
%

In the third stage, \system{} uses the highest-ranked Joint Coordinate System candidate to construct the keyframe sequence.
Using the preceding outputs and the rendering code, the Keyframes Agent specifies the keyframes according to \autoref{sec:transition-space-keyframe-sequences}.
These specifications form the Keyframe Table, with one row per keyframe in sequence order.
The Agent also records which elements persist across adjacent keyframes and which visual properties must remain unchanged.
%

The Keyframe Renderer Agent renders the keyframes according to the Keyframe Table.
To support smooth playback, it selects SVG, Canvas, or raster rendering for each layer based on the number of marks and how they are updated.
It preserves identifiers for elements that persist across keyframes, allowing the Transition Agent to locate the same elements later.
It also checks that the first and last keyframes preserve the original panels' contents and internal arrangement, allowing only overall translation and uniform scaling.
The rendered keyframes are then presented to the user for review.
%

\subsection{Transition Generation}
\label{sec:transition-generation}

Once the user confirms the keyframes, the Transition Agent processes adjacent keyframe pairs in sequence.
Using Shared Data and its associated Transforms, the Keyframe Table, and the rendered keyframes, it determines element correspondences and changes during the transition.
%

Following prior authoring systems, \system{} describes object changes with animation rules such as matching, entering, exiting, merging, and splitting, and organizes these rules into transition steps~\cite{thompson2021dataanimator,kim2021gemini}.
The Transition Agent selects animation rules according to the available correspondence evidence (\autoref{fig:transition-correspondence}(a--d)).
For a confirmed record-level correspondence, the Agent preserves the mark identity and interpolates changes in position and visual attributes.
When a declared Transform connects the results shown in the two keyframes, the Agent maps \(n\) input marks to \(m\) output marks according to the Transform, where \(n\) and \(m\) may each be one or greater.
When a shared level value is the only relation, the Agent treats all marks representing that value in each keyframe as a group and establishes correspondence between the two groups.
When no such relation can be determined, the source marks exit and the target marks enter independently.
%

The Transition Agent determines the order and timing of changes and generates the animation between adjacent keyframes.
The Animation Refiner Agent then reviews the complete sequence for discontinuities in motion and abrupt transitions between stages.
It refines element movement and timing to improve continuity, pacing, and visual clarity.
%

\subsection{Implementation}
\label{sec:system-implementation-runtime}

The \system{} frontend is built with React and TypeScript, and its server is implemented in TypeScript on Node.js.
The Agents use GPT-5.6 Sol through the Codex CLI.
Each Agent in the generation pipeline receives a dedicated prompt, source materials, and validated outputs from preceding stages rather than prior conversation history.
For structured JSON outputs, validation scripts check schema compliance, references to preceding outputs, and compliance with task rules before passing the results to the next Agent.
\system{} also executes the rendering code in headless Chromium to verify that the keyframes render correctly and that each transition starts and ends at the corresponding keyframes.
%

%

\section{Evaluation}

We evaluate our approach through three studies.
Case studies examine how \system{} constructs transitions for scientific figures with different chart types in different fields.
A user study with scientific authors assesses the \system{} authoring workflow, and an expert evaluation examines the generated transitions.
%

\textbf{Panel Reconstruction.}
To prepare the materials for these studies, we prioritized papers that provided source data and plotting code and used these materials to reproduce the selected panels.
When materials were unavailable, or full reproduction required large datasets or specialized processing pipelines, we recreated the affected chart elements using approximate data derived from their positions in the published panels.
Each selected panel was recreated as a data-driven SVG and iteratively compared with the published image at its original resolution.
%

\subsection{Case Studies}

We discuss two representative cases in detail.
\autoref{fig:case}(a) shows an additional transition from a histogram to a higher-order network.
%

\subsubsection{From Weekly Temporal Patterns to Spatial Partitions}

The case in \autoref{fig:teaser}(a) examines weekly patterns of crime incidents and traffic collisions across time and space~\cite{prietocuriel2021heartbeat}.
For each event type, seven area curves show the weekly patterns of categories A--G.
The spatial partition map uses the same category labels and colors to connect each weekly pattern to the corresponding group of regions across the city.
%

The Figure Table records the Transforms and Encodings of the weekly curves and spatial regions, including the category field shared by the two panels.
The Layout is
\((\mathrm{area}+\mathrm{map})\times\mathrm{Event}(2)\).
Within each event type, the Transition Pairing Agent uses the \(+\) operator to recommend an
\(\mathrm{area}\rightarrow\mathrm{map}\) pair (\autoref{sec:transition-pairing}).
We illustrate the recommended pair for crime incidents.
%

\begin{figure*}[!t]
  \centering
  \includegraphics[width=\textwidth, alt={Three side-by-side panels show seven questionnaire rating distributions, Raw NASA-TLX workload scores, and FigFlux SUS scores for 16 participants.}]{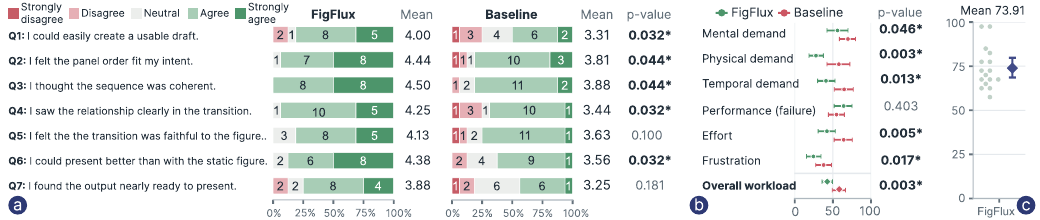}
  \caption{User study results (\(n=16\)). (a) Five-point questionnaire ratings, with segment labels showing participant counts. (b) Mean Raw NASA-TLX scores. (c) Individual \system{} SUS scores (dots) and their mean (diamond). Error bars indicate pointwise 95\% bootstrap CIs for means. Higher scores in (a,c) and lower scores in (b) are favorable. Asterisks indicate BH-adjusted $p<.05$.}
  \label{fig:user-study}
\end{figure*}

The two panels share category values A--G.
Shared Data retains the source rows supporting each category separately for the two panels.
Because the curves' position encodings and the map's geographic layout cannot be preserved in a single 2D coordinate system, \system{} constructs a 3D Joint Coordinate System that preserves panel planes.
The keyframe in \autoref{fig:teaser}(\(a_1\)) represents Combined Data with all seven temporal curves and all spatial regions.
The regions are arranged on separate planes by category and aligned with their corresponding curves.
During the transition, \system{} maintains correspondence between each curve and the group of regions sharing its category (\autoref{fig:transition-correspondence}(c)).
%

\subsubsection{Observed and Perceived Classroom Popularity}

\autoref{fig:case}(b)~\cite{kovarik2025perception} compares participants' observed and perceived popularity across 11 classroom networks.
The networks show the strong-tie relations among participants, while the accompanying histograms and bubble scatter plots compare observed and perceived measures of popularity and centrality.
Observed popularity is the number of classmates who identify a participant as a strong tie, and perceived popularity is the participant's estimate of that number.
The histogram compares the two distributions.
The bubble scatter plot places observed popularity on the \(x\)-axis and perceived popularity on the \(y\)-axis, with bubble size indicating the number of participants sharing the same pair of values.
%

The Figure Table records the Transforms and Encodings used to construct the network, histogram, and bubble scatter panels from the participant and relation data.
The Layout is
\((\mathrm{network}\times\mathrm{Class}(11))+((\mathrm{histogram}+\mathrm{scatter})\times\mathrm{Metric}(2))\), where the two metrics are popularity and centrality.
We illustrate the transition from the popularity histogram to the corresponding bubble scatter plot.
%

Both panels summarize the same participants.
Shared Data retains their source records separately for each panel.
The keyframe in \autoref{fig:case}(\(b_1\)) represents Combined Data with one mark per participant, using observed popularity for \(x\) and perceived popularity for \(y\).
The observed and perceived distributions remain visible as marginal histograms at the top and right, respectively, with bar lengths encoding frequency.
This 2D Joint Coordinate System reuses the target scatter plot's axes (\autoref{fig:transition-space}(\(b_1\))).
\system{} expands histogram bars into participant marks using the source records and the binning and aggregation Transforms recorded in the Figure Table.
It then merges marks with the same pair of values into bubbles (\autoref{fig:transition-correspondence}(b)).
%

\subsection{User Study}

Our within-participant study compared \system{} with a baseline of familiar tools chosen by participants (e.g., PowerPoint and ChatGPT) for animating multi-panel scientific figures.
Each participant completed one task per condition using different figures, with condition order and figure assignment counterbalanced within assigned field groups.
%

\textbf{Participants.} We recruited 16 domain experts from the eight Nature Portfolio subject areas used in our corpus.
All had authored at least one peer-reviewed paper containing a multi-panel scientific figure and had designed and presented that figure.
Participants reported 2--5 years ($n=13$) or 6--10 years ($n=3$) of research experience.
All participants had experience presenting multi-panel scientific figures, with 13 doing so often or almost always.
Participants generally reported familiarity with PowerPoint and generative AI tools.
They received the equivalent of approximately US\$12 per hour and provided informed consent before the study, which our institutional review board approved.
%

\textbf{Study Materials.} We selected 16 published multi-panel scientific figures, two per field, each containing at least one panel pair with a clear scientific relationship.
All candidate transitions were generated in advance to avoid interrupting authoring.
For the baseline condition, we provided the paper, data, and all other available source materials.
%

\textbf{Procedure.} Participants received their two assigned figures, captions, abstracts, and relevant paper excerpts 24--48 hours in advance to review the scientific content.
Before the \system{} task, we demonstrated its main functions using a separate example figure.
Participants were asked to think aloud during both tasks.
Participants could choose which panels to include, but the two specified panels had to appear consecutively in the assigned order.
With \system{}, they reviewed and previewed candidate transitions, selected those to use, and adjusted content and timing as needed.
%
Participants ended each task when they felt the transitions supported a clear explanation, then used them to present the figure and completed a post-task questionnaire.
A semi-structured interview after both tasks compared the two workflows.
%

\subsubsection{Quantitative Results}

Post-task ratings for the two workflows are summarized in \autoref{fig:user-study}.
We used two-sided Wilcoxon signed-rank tests to compare participants' ratings between the \system{} and baseline conditions.
Benjamini--Hochberg correction was applied separately to the seven selected questionnaire items and the seven workload measures (six dimensions and the overall score).
All reported $p$-values are adjusted.
%

\textbf{Authoring support and output quality.}
\system{} received significantly higher ratings for ease of draft creation, alignment of panel order with participants' intended explanation, and coherence of the explanatory sequence (Q1--Q3, all $p<.05$).
It also received higher ratings for clarity of panel relationships and support for presenting the figure compared with the static figure alone (Q4 and Q6, both $p<.05$).
Q5 and Q7 mean ratings were numerically higher with \system{}, but neither difference was statistically significant ($p=.100$ and $.181$, respectively).
%

\textbf{Workload.}
We used NASA-TLX~\cite{hart1988nasatlx} to assess workload, computing the Raw TLX score as the unweighted mean of its six dimensions.
The overall score was significantly lower with \system{} than with the baseline (42.86 vs.\ 58.44, $p=.003$).
Mental, physical, and temporal demand, effort, and frustration were also significantly lower (all $p<.05$).
Performance did not differ significantly ($p=.403$).
%

\textbf{Usability.}
\system{} received a mean SUS score of 73.91 (SD = 11.80), consistent with ``good'' perceived usability based on Bangor et al.'s adjective ratings~\cite{bangor2009sus}.
SUS was administered only for \system{}.
%

\subsubsection{Qualitative Feedback}

\textbf{Explaining data relationships between panels.}
Participants reported that \system{} helped them explain data relationships between panels.
P10 noted that PowerPoint supported detailed explanations of individual charts, while \system{} more clearly showed the spatiotemporal correspondence between them.
P16 also felt that the animation clearly showed how the two panels were related.
P7 and P11 said that the intermediate keyframe helped explain the connection between the two panels.
P11 noted that the transition from monthly to annual information followed the intended order of explanation.
P2 and P9 similarly felt that showing these relationships through animated transitions reduced the need for verbal explanation.
Reflecting on their earlier explanations, P2 worried that verbal descriptions might not have conveyed their intended meaning to the audience.
When using transitions, P12 explained each panel's findings and also how the panels were connected.
P12 also realized that audiences in previous presentations might not have understood these connections, so an explanation was needed.
%

\textbf{A starting point for presenting figures.}
In the baseline task, P12 spent a long time formulating a prompt for an AI assistant.
Four participants (25\%) did not submit any requests to the Agent.
P6 and P7 reported that the default transition content was sufficient for presenting the figure.
P6 noted that \system{} reduced manual work in preparing visual materials, allowing them to focus on presenting the figure and follow the animation's sequence.
P12 noted that \system{}'s generated keyframes broke the content into presentation steps and structured the explanation, reducing the need to plan these steps from scratch.

\textbf{Editing transitions with the Agent.}
Participants who wanted further changes used \system{}'s Agent to edit the generated transitions.
P9 asked for gradual changes between keyframes to avoid abrupt jumps and said the second revision met expectations.
P4 kept the red basemap unchanged during the transition to make geographic locations easier to identify.
P13 was unsure whether the request was clear but said the final revision met expectations.
P13 reported similar concerns when using other AI tools and did not consider this a problem specific to \system{}.
P10 wanted to draw a selection box on a keyframe before describing the desired change to the Agent.
P10 said text alone could make it difficult to specify the exact part to edit.
P14 valued keyframe-by-keyframe playback, pause controls, and selecting which figure parts to connect.
In the baseline task, P14 wanted a clean map as a background and had to manually delete the other elements one by one.
%

\subsection{Expert Evaluation}

Four visualization experts (E1--E4, 5--8 years of experience) evaluated animations for presenting panel relationships in academic scenarios.
We compared participants' final \system{} and baseline outputs with AI-only results generated using the same model, a common prompt, and the same materials, without \system{} guidance.
All three conditions targeted the same panel pair and direction for each of the 16 user-study figures.
Each expert reviewed eight figures, one per field, with two experts per figure.
Experts viewed anonymized results in balanced order and rated semantic faithfulness, relationship understandability, focused attention, transition coherence, and visual appeal on seven-point scales.
They could consult the figures and relationship descriptions or replay results.
After locking ratings, we conducted semi-structured interviews and revealed production methods after the first question.
%

We summarized the 32 ratings per condition and dimension using means and standard deviations.
\system{} received higher mean ratings across the five dimensions ($M=5.94$--$6.22$, $SD=0.82$--$0.95$) than baseline ($M=3.09$--$4.31$, $SD=1.54$--$1.89$) and AI-only ($M=4.47$--$4.88$, $SD=1.32$--$1.65$).
The same pattern held for each expert's mean ratings across their assigned figures, although agreement on individual clip ratings was limited (ordinal Krippendorff's $\alpha=0.38$--$0.58$).
%

%

Experts valued how the Joint Coordinate System made data relationships between panels visible.
E3 attributed the coherence of the transition to retaining one panel's encoding while introducing the other's in the same view.
E3 also noted that even when two panels showed the same data, their correspondence could be hard to see if their contents never appeared together.
E2 similarly appreciated seeing shared data attributes mapped across panels.
Making these mappings visible, E2 felt, could reduce the need to consult the captions or main text to understand how the data corresponded.
E4 envisioned readers first seeing how the panels relate and then consulting the caption for details.
Moreover, E3 noted, the original panels were already complex and explaining their relationship in words would be no simpler.
Presenting their content in smaller parts could instead make the relationship easier to explain.
%

Experts also appreciated the use of 3D.
E2 particularly liked the 3D effects and the data mappings.
E4 felt that 3D provided more space to spread out visual elements such as bars, reducing overlap and offering more ways to arrange and animate them.
Compared with the baseline results, he found it easier to see where the same element was located in each of the two panels.
E3 also preferred this 3D presentation even for simple charts, as it showed content from both panels at once.
%

E4 noted that some baseline results simply placed visual elements on top of one another.
E3 thought AI-only worked well in some cases.
After viewing several results, however, he found that the animations changed little across cases, even when the data relationships differed.
He also noted that some elements appeared without a transition linking them to the previous panel.
E1 appreciated \system{}'s focus on relationships between panels and felt that AI still needed expert guidance to help authors tell clear data stories.
%

\section{Discussion}
\label{sec:discussion}

\textbf{Exploring the Boundaries of AI in Visualization.}
Large language models enable us to generate intermediate keyframes and transitions that reveal data relationships between panels in scientific figures.
Joint Coordinate Systems bring together data and visual elements from separate panels to show how they relate.
In particular, a 3D Joint Coordinate System allows charts with different coordinate structures to retain their spatial layouts in a shared view, while animated transitions reveal connections between their data.
Creating these keyframes and transitions requires tracing data relationships, arranging visual elements, and writing animation code for each figure.
In \system{}, large language models help carry out this work, guided by the representations and rules defined in Figure Space and Transition Space.
%

Some participants were surprised by the Joint Coordinate Systems, noting they had not considered presenting data this way.
Some also felt that the 3D views could help their work make a strong impression on audiences.
These responses encourage us to explore the boundaries of AI in visualization.
This means asking which design ideas become feasible as models evolve and what new representations and methods are needed to realize them.
Through this process, we hope to translate AI advances into more useful ways to understand data.
%

\textbf{Generalizability.}
Our method could also support understanding data relationships across charts in a wider range of applications.
E2 envisioned using the Joint Coordinate System in multiview visual analytics systems to identify and highlight related data from different views.
E1 suggested financial presentations, and E3 highlighted applications on mobile devices.
Showing data correspondences directly could make comparisons across charts easier.
Specifically, in visual analytics systems, future work could explore generating transitions based on the data users select and the views they are comparing, helping them follow how those data are represented in each view.
On mobile devices, transitions could reduce switching between charts, with layouts adapted to keep marks and labels legible.
We also envision applications in education, where interactive papers and teaching materials could let students pause or replay transitions to examine how data correspond across charts.
%

\textbf{Limitations.}
First, our method relies on source data and rendering code.
We reconstructed the charts as closely as possible from the available materials, but those based on approximate data may not exactly match the originals.
Second, although our method automates animation work that can be time-consuming even for experts, generation still takes time.
Users can work on other tasks while the system generates the transitions.
Future improvements to models and the workflow could shorten transition generation time.
Third, our system currently prioritizes 2D Joint Coordinate Systems when the panels' visual encodings permit.
In future work, we aim to support customizing these coordinate arrangements based on users' needs.

%

\section{Conclusion and Future Work}
\label{sec:conclusion}

We developed Figure Space and Transition Space for representing scientific figures and designing keyframes that explain data relationships across their panels.
Building on these spaces, \system{} guides AI agents in generating animated transitions that researchers can review and revise.
Case studies, a study with 16 scientific authors, and expert evaluation support its usefulness for authoring animated explanations of scientific figures.
In future work, we plan to extend \system{} to a broader range of visualizations.
We also aim to help authors combine animated transitions with explanatory text and highlights to better support their communication goals.
%

\interlinepenalty=10000 
\bibliographystyle{IEEEtran}
\bibliography{template}


\end{document}